\documentclass[a4paper,11pt]{article}
\usepackage{pos}
\usepackage{bbm}
\usepackage{hyperref} 
\usepackage{caption}    
\usepackage{amsmath}
\usepackage{float} 
\usepackage{graphics}
\usepackage{graphicx} 
\usepackage{multirow}
\usepackage{array}

\newcommand{\beq}{\begin{eqnarray}}
\newcommand{\eeq}{\end{eqnarray}}

\title{Radion dynamics in  multibrane Randall-Sundrum model}

\author*[a]{Haiying Cai} 

\affiliation[a]{Department of Physics, Korea University, Seoul 136-713, Korea}

\emailAdd{hcai@korea.ac.kr}

\abstract{The radion  in the Randall-Sundrum model is stabilized by the back reaction of a bulk scalar field with its VEV depending on the fifth dimensional coordinate. We studied the radion dynamics in an extended scenario, where  intermediate branes exist between the UV and IR branes. Our analysis proves  that the formalism of EFT delivers the same equations of motion as the linearized  Einstein equation with all junction conditions satisfied.  The relic of 5d diffeomorphism  is broken after including the Goldberger-Wise stabilization and a unique radion field is conjectured as legitimate in the RS metric perturbation. }

\FullConference{%
  41st International Conference on High Energy physics - ICHEP2022\\
  6-13 July, 2022\\
  Bologna, Italy
}

\begin{document}
\maketitle

\section{5d warped model} 
The Randall-Sundrum (RS) model compacted in a slice of   anti–de-Sitter (AdS)  space with 2 branes can naturally illustrate the emerging  of  TeV energy scale and  the weakness of gravity~\cite{Randall:1999ee}. The multibrane RS model  is an  appealing extension since a new energy scale between the Planck  and TeV ones is plausible for  phenomenology study~\cite{Kogan:2001qx}.  The  $N $-brane ($N \in$ even integer) extension of RS model can be achieved given that the cosmology constants $\Lambda_\alpha$, $\alpha = 1, \cdots \frac{N}{2}$ are different in each subregion of an AdS$_5$ space.  We start with the generic five dimensional action with an orbifold symmetry for the graviton coupling to a single bulk scalar field~\cite{DeWolfe:1999cp, Csaki:2000zn}: 
\beq
&& S =  S_{EH} + S_m =  -  \frac{1}{2 \kappa^2} \int d^5 x \sqrt{g} \, {\cal R} + 
\int d^5 x \sqrt{g}  \mathcal{L}_m  \,, \label{Act} 
\\ 
&& \mathcal{L}_m =  \frac{1}{2} g^{I J}  \partial_I \phi \partial_J \phi
-  V(\phi) - \frac{1}{\sqrt{-g_{55}}}\sum_{i=1}^N \lambda_{i}(\phi) \delta(y - y_i)  
\end{eqnarray}
where the first term in Eq.(\ref{Act}) is the Einstein-Hilbert action, and the second one is the matter action. For simplicity we refer all objects located at $y_i$,$i = 1, \cdots N$ as branes like in the original RS model. With appropriate bulk and brane potentials $V(\phi)$ and $\lambda_i$, the scalar $\phi = \phi_0(y) + \varphi(x, y)$ develops a $y$-dependent VEV  that reacts back on the metric. As a result,  the radion field  will obtain a mass via the Goldberger-Wise (GW) mechanism~\cite{Goldberger:1999uk}.  The line element in an $N$-brane model decoupling the graviton and radion is~\cite{Kogan:2001qx, Cai:2021mrw}:
\beq
d s^2  &=&   e^{-2 A (y) - 2 F(x, y) }\left[ \eta_{\mu \nu} + 2 \epsilon(y) \partial_{\mu} \partial_{\nu} f (x) \right. \nonumber \\ 
&+& \left. h_{\mu \nu}(x,y)\right] d x^\mu d x^\nu - \left[1+ G(x, y) \right]^2 dy^2  \,, \label{metric}
\eeq
where the $\epsilon(y)$ is added as a radion perturbation in addition to $F$ and $G$. By varying the 5d action with respect to the metric $g_{IJ}$, one can derive the Einstein equation: $\mathcal{R}_{IJ}  = \kappa^2 \left(T_{IJ} -\frac{1}{3} g_{IJ} T^m_m\right) = \kappa^2 \tilde{T}_{IJ}$, with the energy-momentum tensor defined as $T_{IJ} = 2 \delta (\sqrt{g} \mathcal{L}_m)/ (\sqrt{g} \delta g^{IJ})$. While minimizing $S_m$ with respect to  $\varphi$ gives  the scalar EOM, that at the linear order is modified by a term $\phi'_0 \epsilon' \Box f(x)$  in  our metric~\cite{Cai:2022geu}. But  the solutions for background metric and scalar VEV $\phi_0$  are the same as in RS1 and can be written in terms of a single super-potential $W(\phi)$ as~\cite{DeWolfe:1999cp, Behrndt:1999kz}:
\beq
\phi_{0} ' = \frac{1}{2} \frac{\partial W}{\partial \phi} \,, \quad 
A'=\frac{\kappa^2}{6} W(\phi_0) \,, \quad V(\phi) = \frac{1}{8}
\left[\frac{\partial W(\phi)}{\partial \phi}\right]^2 - \frac{\kappa^2}{6} W(\phi)^2
\eeq
We are going to investigate the graviton-scalar system in the formalism of effective Lagrangian and  practice the variation principle to a specific perturbation field. In fact  the scalar EOM has to be derived in this approach as it is not contained in the Einstein equation. Expanding the 5d action Eq.(\ref{Act}) till the quadratic order,  we obtain the effective Lagrangian~\cite{Cai:2022geu}:
\beq  
\mathcal{L}_{eff} & = & \int dy  \Big\{\frac{e^{-2A}}{2\kappa^2} \Big[  \frac{e^{-2A}}{4} \left[(\partial_5 h)^2 - \partial_5 h_{\mu \nu} \, \partial_5 h^{\mu \nu} \right]  - {\mathcal L}_{FP} \Big]\nonumber \\& + &  {\mathcal L}_{mix} \, +  \, \mathcal{L}_{r-kin} - \frac{e^{-4A}}{2}  \mathcal{L}_{5m}  \Big \}  \, \label{eff}
\eeq
where the terms in the first line are for the graviton with:
\beq
{\mathcal L}_{FP} = \frac{1}{2} \partial_\nu h_{\mu \alpha} \, \partial^\alpha    
h^{\mu \nu} -   \frac{1}{4} \partial_\mu h_{\alpha \beta} \, \partial^\mu h^{\alpha \beta}    
- \frac{1}{2} \partial_\alpha h \, \partial_\beta h^{\alpha \beta} +\frac{1}{4}    
\partial_\alpha h \, \partial^\alpha h \,
\eeq
The second line in Eq.(\ref{eff}) contains the graviton-radion mixing term along with the kinetic and non-kinetic terms of radion:
\beq
\mathcal{L}_{mix} &=&  -\frac{ e^{-2A}}{2 \kappa^2}\left[ \left[ G- 2 F - e^{2A}  \partial_5 \left( \epsilon'  f(x) e^{-4A}\right) \right]   \left( \partial_{\mu} \partial_\nu h^{\mu \nu} - \Box \, h\right) \right. \nonumber \\ &+ &\left. 3 e^{- 2A} \left[ F'  -A' G  - \frac{\kappa^2}{3} \phi'_0 \varphi \right] \partial_5  h  \right]
\, \label{mix} \\
 \mathcal{L}_{r-kin} &=&  \frac{1}{2} \int dy e^{-2A} \kappa^2  \partial_\mu \varphi \partial^\mu \varphi  -\frac{6}{\kappa^2} \Big[\partial_\mu F \partial^\mu \left( F- G \right)  \nonumber \\ &-&  e^{-2A} \epsilon'  \partial_\mu \left[ F'  -A' G  - \frac{\kappa^2}{3} \phi'_0 \varphi \right]  \partial^\mu f(x) \Big] 
 \\
\mathcal{L}_{5m} & = &- \frac{12}{\kappa^2}  \Big[ F'^2 +  G^2 A'^2 - 2 F' G A'   \Big]  +  \varphi'^2 + G^2  \phi'^2_0 -  2 (G+ 4F) \phi'_0 \varphi' \nonumber \\ &+&   \left[ 2 (G- 4F)  \frac{\partial V}{\partial \phi_0} \varphi  +   \frac{\partial^2 V}{\partial \phi_0^2} \varphi^2 \right] 
- \sum_{i=1}^N \Big[  8  \frac{\partial \lambda_i}{ \partial \phi_0} F \varphi  -  \frac{\partial^2 \lambda_i}{\partial \phi_0^2}\varphi^2  \Big] \delta(y-y_i) 
\eeq
Requiring  no-mixing between the graviton and radion, one can immediately derive two orthogonal conditions from Eq.(\ref{mix}):
 \beq
&& F'  - A' G  -  \frac{\kappa^2}{3} \phi'_0 \varphi  =0  \,,    \\
&& G - 2 F  -  e^{-2A} \left[ \epsilon'' - 4 A'\epsilon'\right] f(x)  =0 \,.  \label{orth}  
\eeq  
\begin{figure}
	\centering 	
	{\includegraphics[height=3.2 cm,  
		width=6.0 cm]{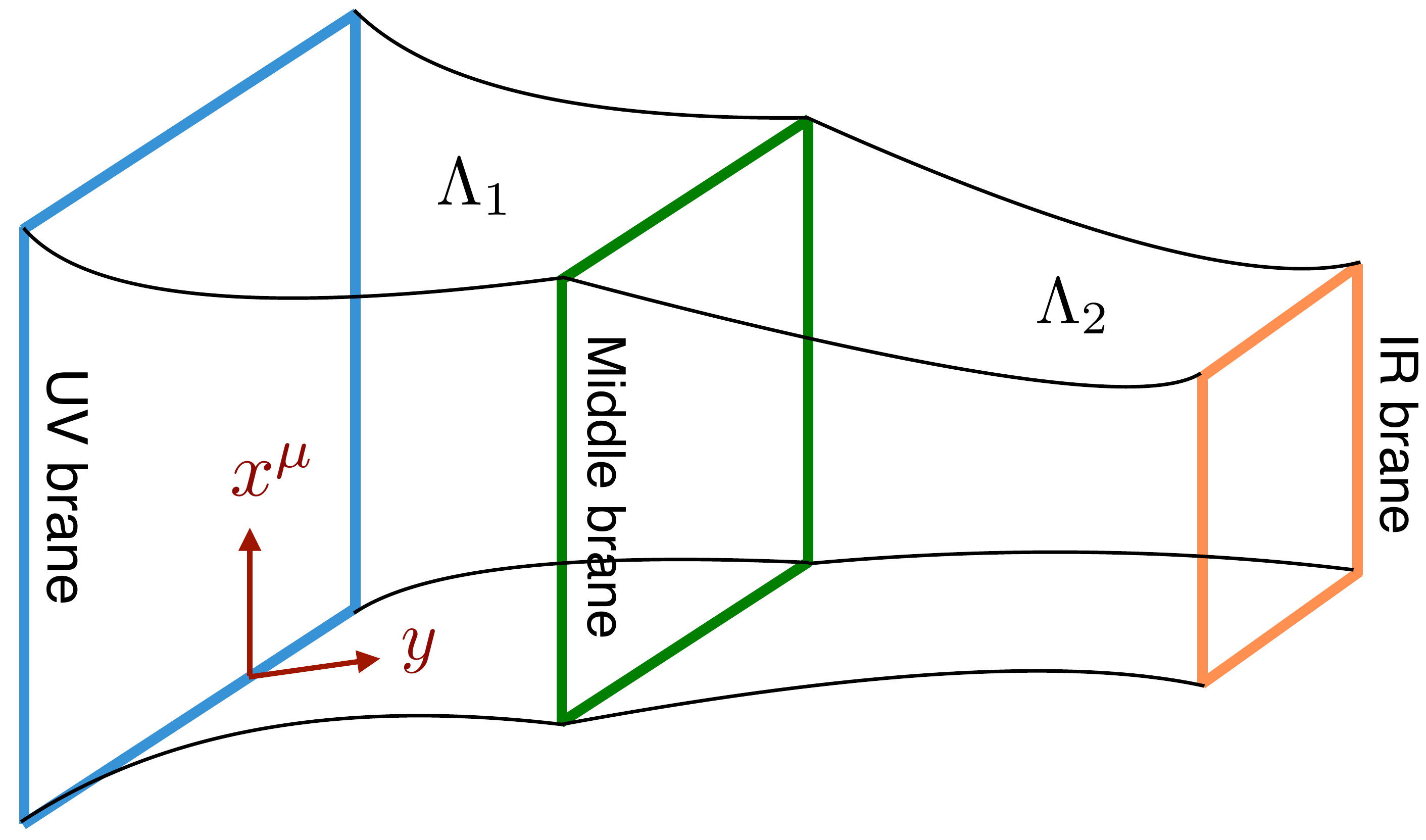}} 
	\caption{The $4$-brane  RS model  visualized in the range of $0 \leq y \leq L$. An identical part exits in $-L \leq y < 0 $ due to the orbifold symmetry.} \label{fig: 4ibrane}
\end{figure} 
By varying Eq.(\ref{eff}) with respect to $F, G$ and $\varphi$, one can obtain 3 EOMs. We find that one exact correspondence can be established between 2 EOMs and the linearized Einstein equations, i.e.
\beq
\frac{\delta \mathcal{L}_{eff}}{\delta G}=0  &\Rightarrow &\frac{1}{\kappa^2}\left[e^{2 A} \mathcal{R}_{\mu \nu}/\eta_{\mu \nu} + \mathcal{R}_{55} \right] =  \left[e^{2 A} \tilde{T}_{\mu \nu}/\eta_{\mu \nu} +\tilde{T}_{55} \right]  \,, \label{E1}
\\
\frac{\delta \mathcal{L}_{eff}}{\delta F}=0  &\Rightarrow & \frac{1}{2\kappa^2}\left[2 e^{2 A} \mathcal{R}_{\mu \nu}/\eta_{\mu \nu} - \mathcal{R}_{55} \right] =  \left[e^{2 A} \tilde{T}_{\mu \nu}/\eta_{\mu \nu} -\frac{1}{2}\tilde{T}_{55} \right]  \,. \label{E2}
\eeq
The 3rd one $\frac{\delta \mathcal{L}_{eff}}{\delta \varphi}=0$ gives the scalar EOM:
\beq
 && (G' + 4F' ) \phi'_0 +   4 A' \varphi'   + \sum_{i=1}^{N} \left (  \frac{\partial \lambda_i}{ \partial \phi_0} G + \frac{\partial^2 \lambda_i}{ \partial \phi_0^2} \varphi  \right) \delta(y-y_i) \nonumber \\
&=& \varphi''  - \left( 2   \frac{\partial V}{\partial \phi_0} G  + \frac{\partial^2 V}{\partial \phi_0^2} \varphi \right) -  \Box  \Big( \varphi  e^{2A} -\phi'_0 \epsilon' f(x) \Big) \,.  \label{scalar}
\eeq
One can prove that Eq.(\ref{E1}-\ref{scalar}) are correlated and  a single EOM is independent~\cite{Cai:2022geu}. With some algebra, Eq.(\ref{E1}) can be recasted into the familiar form~\cite{Cai:2021mrw}: 
\beq
&& 3\left( F'' - A' G' \right) + 3 \left[ F e^{2 A} - A' \epsilon' (y) \right] \Box f (x)  \nonumber \\
&=& 2 \kappa^2 \phi_0' \varphi' + \frac{\kappa^2}{3} \sum_{i=1}^{N} \left[ 3 \lambda_i (\phi_0)  G   +  3 \frac{\partial \lambda_i}{\partial \phi} \varphi \right] \delta(y - y_i) 
 \label {eom}
 \eeq
In fact  only one EOM plus two orthogonal conditions are independent for 4 radion fields $F, G$, $\varphi$ and $\epsilon$, and  this implies  that one perturbation is a gauge fixing. 

\section{Breaking of  5d  diffeomorphism}
We will first review the 5d diffeomorphism  that is  an infinitesimal coordinate  transformation $X^I \to X^I + \xi^I(X)$ which can keep $S_{EH}$ (not the full 5d action) to be invariant.  As a result, the metric  transforms accordingly:
\beq 
\delta g_{IJ} = - \xi^K \, \partial_K  g_{IJ}^{(0)} - \partial_I \xi^K \,  g_{K J}^{(0)} - \partial_J \xi^K \, g_{IK}^{(0)} \,. \label{dG}
\eeq
To retain the metric in its original structure after a field redefinition, $\xi^I(X)$ is constrained to  be of  the specific form~\cite{Kogan:2001qx}:
\beq
\xi^\mu(x,y) &=& \hat{\xi}^\mu(x)  + \, \eta^{\mu \nu}  \partial_\nu   \zeta(x, y) 
\nonumber \\   \xi^5(x,y) &=& e^{-2A} \, \zeta^\prime(x, y) \,. \label{xi}
\eeq
with  $\zeta^\prime(x, y)|_{y = y_i}=0$. Hence the metric perturbations transform as:
\beq
&& \delta h_{\mu \nu} = - \partial_{\mu} \hat{\xi}_{\nu} - \partial_\nu \hat{\xi}_\mu \nonumber\\
&&\delta F = - A'  \zeta'  e^{-2A} \nonumber \\
&&\delta G =  - \left( \zeta'' - 2 A' \zeta' \right)  e^{-2A} \nonumber\\
&& \delta \epsilon f(x) =- \zeta   \, \label{transform}
\eeq
where the last ansatz indicates that $\epsilon f(x)$ plays the role of gauge fixing and can be removed by setting $\zeta = \epsilon f(x)$.
However the situation becomes different  after the radion stabilization. Inspecting Eq.(\ref{E1}-\ref{scalar}), we can find that a field redefinition below can remove the $\epsilon$ dependence in all 3 EOMs:
\beq 
\tilde{F} &=& F - A' \epsilon' f(x) e^{-2A} \nonumber  \\
\tilde{G} &=& G - \left( \epsilon'' - 2 A' \epsilon' \right) f(x) e^{-2A}  \nonumber  \\
\tilde{\varphi} &=& \varphi - \phi'_0 \epsilon' f(x) e^{-2A} \, \label{zeta}     
\eeq
where the first two  are  the same as in the diffeomorphism (see Eq.(\ref{transform}))  and the last one is for  the GW scalar.  As an invariant scalar, $\delta \varphi =0 $ forces $ \phi'_0 \, \epsilon' =0$ otherwise the 4d Poincar\'e symmetry will be broken and  ambiguity will enter in the radion kinetic term. Since we pursue a solution with a stabilized radion,  the correct option is to let  $\phi'_0 \neq 0$ and  $\epsilon' =0$~\cite{Cai:2022geu}. This signals that the 5d $\zeta$-symmetry is actually broken.

\section{Stabilization and cosmology implication} 
The solvable radion EOM in a multibrane RS model is Eq.(\ref{eom}) gauged with $\epsilon' = 0$ and $G = 2 F$. We consider the simplest case $N=4$,  and generalize the GW mechanism to the $4$-brane model  by choosing the following superpotential,
\beq
W(\phi) = \begin{cases} \frac{6 k_1}{\kappa^2} - u \phi^2 \,, &  0 < y < r  \\[0.3cm]     
\frac{6 k_2 }{\kappa^2} - u  \phi^2  \,,  & r < y < L  
\end{cases}    \, \label{pot}
\eeq
To solve the eigenstate problem,  the stiff  potentials  will be imposed at the UV and IR branes similar to RS1 such that $\varphi|_{y=\{0, L\}} =0$. While at the intermediate brane $y=r$,  we  use the junction condition $\left[ F' \right]_r  =  2  \left[A'\right]_r F(r) $. Then for $l = \frac{ \kappa \phi_0}{\sqrt{2}}|_{y=0} \ll 1$,  the radion mass turns out to be $m^2 \simeq \frac{ 4 u^2  (2 k_2 + u)l^2}{3 k_2} e^{-2 \left[(k_2+ u) L + ( k_1 - k_2) r \right]}$, that is below the cut off scale of IR brane. We can briefly discuss the cosmological evolution in the multibrane RS model by taking the metric to be time dependent~\cite{Binetruy:1999ut, Csaki:1999mp}: 
\beq
&& ds^2 \,= \, n(t, y)^2 dt^2 - a(t, y)^2  dx^2 - b(t, y)^2 dy^2  
\nonumber \\
&& a(t, y)  = a_0(t) e^{-A} (1+ \delta a) \,, ~~  n(t, y) = e^{-A} (1+ \delta n)  \nonumber \\
&& b(t, y)  = 1+ \delta b 
\eeq
with the perturbations $(\delta a, \delta n,\delta b)$  caused by adding the matter densities.  Averaging the Einstein Equation $G_{55} = \kappa^2 T_{55}$ with respect to $y=r$, one can derive~\cite{Cai:2021mrw}:
\beq
 \left( \frac{\dot a_0}{a_0} \right)^2  + \frac{\ddot a_0}{a_0} = \frac{  \kappa^2  e^{-2 A}  }{3 }  \left[\frac{k_1 k_2   }{k_1 - k_2}  \left( \rho - 3 p \right)  +  \phi'^2_0   \delta b  \right] \,, \label{FRW}
\eeq
where $\rho$ and $p$ are the matter density and pressure at  the $y=r$ brane. Because the intermediate brane essentially is a  spacetime defect, one expects that $\delta b(r)$ is  fixed by the initial condition.  And Eq.(\ref{FRW}) is a constraint that a multibrane system needs to observe.

\section{Conclusion}
We examine the radion dynamics in  the multibrane RS model by adding a new radion perturbation $\epsilon \,\partial_\mu \partial_\nu f(x) $ in the metric. Our analysis shows  that  the formalism of effective Lagrangian  is equivalent to the linearized Einstein equation and only a single radion EOM is actually independent no matter with or without  stabilization. The role of $\epsilon$ is  a gauge fixing for the Einstein-Hilbert action  and can be removed by setting $\zeta = \epsilon f(x)$ using the 5d diffeomorphism.  However this 5d symmetry  has to be broken in order to preserve the 4d poincar\'e symmetry in the presence of radion stabilization.  Thus $\epsilon'=0$ holds in the full bulk with its boundary value $\epsilon'(r) =0$,  not possible to create a new radion excitation.  The consequence  of one radion  permitted  in the  5d action Eq.(\ref{Act}) renders the intermediate branes to be nondynamical. The unique radion profile is $F \propto e^{2A} + \mathcal{O}(l^2)$,  satisfying the jump condition  $\left[ F' \right]_{r_a}  =  2  \left[A'\right]_{r_a} F(r_a)$. Therefore  we can simply generalize the GW mechanism into an $N$-brane extension in a  way similar to RS1.


\begin{thebibliography}{99}

\bibitem{Randall:1999ee}
L.~Randall and R.~Sundrum, \emph{{A Large mass hierarchy from a small extra
  dimension}}, \href{https://doi.org/10.1103/PhysRevLett.83.3370}{\emph{Phys.
  Rev. Lett.} {\bfseries 83} (1999) 3370}
  [\href{https://arxiv.org/abs/hep-ph/9905221}{{\ttfamily hep-ph/9905221}}].

\bibitem{Kogan:2001qx}
I.~I. Kogan, S.~Mouslopoulos, A.~Papazoglou and L.~Pilo, \emph{{Radion in
  multibrane world}},
  \href{https://doi.org/10.1016/S0550-3213(02)00009-3}{\emph{Nucl. Phys. B}
  {\bfseries 625} (2002) 179}
  [\href{https://arxiv.org/abs/hep-th/0105255}{{\ttfamily hep-th/0105255}}].

\bibitem{DeWolfe:1999cp}
O.~DeWolfe, D.~Z. Freedman, S.~S. Gubser and A.~Karch, \emph{{Modeling the
  fifth-dimension with scalars and gravity}},
  \href{https://doi.org/10.1103/PhysRevD.62.046008}{\emph{Phys. Rev. D}
  {\bfseries 62} (2000) 046008}
  [\href{https://arxiv.org/abs/hep-th/9909134}{{\ttfamily hep-th/9909134}}].

\bibitem{Csaki:2000zn}
C.~Csaki, M.~L. Graesser and G.~D. Kribs, \emph{{Radion dynamics and
  electroweak physics}},
  \href{https://doi.org/10.1103/PhysRevD.63.065002}{\emph{Phys. Rev. D}
  {\bfseries 63} (2001) 065002}
  [\href{https://arxiv.org/abs/hep-th/0008151}{{\ttfamily hep-th/0008151}}].

\bibitem{Goldberger:1999uk}
W.~D. Goldberger and M.~B. Wise, \emph{{Modulus stabilization with bulk
  fields}}, \href{https://doi.org/10.1103/PhysRevLett.83.4922}{\emph{Phys. Rev.
  Lett.} {\bfseries 83} (1999) 4922}
  [\href{https://arxiv.org/abs/hep-ph/9907447}{{\ttfamily hep-ph/9907447}}].

\bibitem{Cai:2021mrw}
H.~Cai, \emph{{Radion dynamics in the multibrane Randall-Sundrum model}},
  \href{https://doi.org/10.1103/PhysRevD.105.075009}{\emph{Phys. Rev. D}
  {\bfseries 105} (2022) 075009}
  [\href{https://arxiv.org/abs/2109.09681}{{\ttfamily 2109.09681}}].

\bibitem{Cai:2022geu}
H.~Cai, \emph{{Effective Lagrangian and stability analysis in warped space}},
  \href{https://doi.org/10.1007/JHEP09(2022)195}{\emph{JHEP} {\bfseries 09}
  (2022) 195} [\href{https://arxiv.org/abs/2201.04053}{{\ttfamily
  2201.04053}}].

\bibitem{Behrndt:1999kz}
K.~Behrndt and M.~Cvetic, \emph{{Supersymmetric domain wall world from D = 5
  simple gauged supergravity}},
  \href{https://doi.org/10.1016/S0370-2693(00)00095-2}{\emph{Phys. Lett. B}
  {\bfseries 475} (2000) 253}
  [\href{https://arxiv.org/abs/hep-th/9909058}{{\ttfamily hep-th/9909058}}].

\bibitem{Binetruy:1999ut}
P.~Binetruy, C.~Deffayet and D.~Langlois, \emph{{Nonconventional cosmology from
  a brane universe}},
  \href{https://doi.org/10.1016/S0550-3213(99)00696-3}{\emph{Nucl. Phys. B}
  {\bfseries 565} (2000) 269}
  [\href{https://arxiv.org/abs/hep-th/9905012}{{\ttfamily hep-th/9905012}}].

\bibitem{Csaki:1999mp}
C.~Csaki, M.~Graesser, L.~Randall and J.~Terning, \emph{{Cosmology of brane
  models with radion stabilization}},
  \href{https://doi.org/10.1103/PhysRevD.62.045015}{\emph{Phys. Rev. D}
  {\bfseries 62} (2000) 045015}
  [\href{https://arxiv.org/abs/hep-ph/9911406}{{\ttfamily hep-ph/9911406}}].


\end{thebibliography}
\end{document}